\documentclass[12pt]{article}
\usepackage{cy}
\usepackage{rotate}
\usepackage{amssymb,amsthm, amsmath}
\usepackage{graphicx}
\usepackage{multirow}
\usepackage{float}
\usepackage{times}

\newcommand{\vectornorm}[1]{\left|\left|#1\right|\right|}

\title{\textsc{A Markov Random Field Topic Space\\ Model for Document Retrieval}}
\date{}
\author{Scott Hand}

\begin{document}
\bibliographystyle{unsrt}

\maketitle

\abstract

\noindent This paper proposes a novel statistical approach to intelligent document retrieval.  It seeks to offer a more structured and extensible mathematical approach to the term generalization done in the popular Latent Semantic Analysis (LSA) approach to document indexing. A Markov Random Field (MRF) is presented that captures relationships between terms and documents as probabilistic dependence assumptions between random variables.  From there, it uses the MRF-Gibbs equivalence to derive joint probabilities as well as local probabilities for document variables. A parameter learning method is proposed that utilizes rank reduction with singular value decomposition in a matter similar to LSA to reduce dimensionality of document-term relationships to that of a latent topic space.  Experimental results confirm the ability of this approach to effectively and efficiently retrieve documents from substantial data sets.

\tableofcontents \listoffigures

\clubpenalty=9999
\widowpenalty=9999

\section{Introduction}
Research in the field of information retrieval is becoming increasingly important as large sources of data become available and users become accustomed to powerful and flexible ways of processing this information.  It is now accepted that simple data retrieval methods based on naive term matching fail to function effectively for large and varied bodies of data \cite{Deerwester_1990}.  In particular, users are beginning to seek methods of retrieval that examine the meanings of queries rather than the queries themselves.  One promising approach to this, Latent Semantic Analysis (LSA), was proposed by \cite{Furnas_1988} as at attempt to generalize terms into latent topic concepts using linear algebra techniques.  We seek to provide a more structured approach to accomplishing term generalization similar to LSA using a Markov Random Field model.  We believe that this approach has a more solid foundation and provides researchers with a better understanding of the underlying mathematics and potential for extension.

\subsection{Related Work}

\subsubsection{Latent Semantic Analysis} \label{ss:lsa}

Latent Semantic Analysis (LSA) is a method used in information retrieval for smoothing sets of document-term data.  Documents in a large collection are subject to statistical over-specification, as each one only contains a small fraction of the terms despite being relevant with respect to many other terms.  LSA expands upon a vector-space model \cite{Salton_1975} in which documents are represented as row vectors of terms.   A co-occurrence matrix $\mathbf{X}$ representing a collection of documents can be defined as a matrix whose rows are term vectors $\mathbf{T}$ and columns are document vectors $\mathbf{D}$.
\begin{equation*} \mathbf{X} =
\begin{bmatrix}
	 x_{1,1} & \cdots & x_{1,n} \\
	\vdots & \ddots & \vdots \\
	x_{m,1} & \cdots & x_{m,n}
\end{bmatrix}
\end{equation*}
The value $x_{t,d}$ refers to the number of times term $t$ appears in document $d$. This representation is convenient because it allows the similarity of any column vector $\mathbf{d}$ of matrix $\mathbf{X}$ and query vector $\mathbf{q}$ to be calculated as the cosine of the angle between the two vectors using:
\begin{equation} \label{eq:cos_similarity}
\mathrm{cos}(\theta) = \frac
{\mathbf{d}\cdot\mathbf{q}}
{\vectornorm{\mathbf{d}}\vectornorm{\mathbf{q}}}
\end{equation}

One problem with this approach is that, since it relies solely on terms as being independent, it fails to capture the semantic relationship between synonyms and other examples of distinct but related terms.  It also results in poor and uneven recall because it relies on the specific wording of the query, and, without any smoothing, many relevant documents could be missed due to lexical discrepencies.

LSA attempts to generalize terms into a latent topic space by reducing the dimensionality of the co-occurrence matrix.  This is accomplished by first taking a Singular Value Decomposition on the co-occurrence matrix.  This produces three new matrices, $\mathbf{U}$, $\mathbf{S}$, and $\mathbf{V}$ such that $\mathbf{X} = \mathbf{U}\mathbf{S}\mathbf{V}^T$. $\mathbf{U}$ and $\mathbf{V}$ contain orthogonal column vectors while $\mathbf{S}$ is a diagonal matrix.  The diagonal of $\mathbf{S}$ forms a vector of singular values $\mathbf{\sigma}$.
\begin{equation*}
\begin{matrix}
	\mathbf{X} & & \mathbf{U} & & \mathbf{S} & & \mathbf{V}^T \\
	\\
	\begin{bmatrix}
		 x_{1,1} & \cdots & x_{1,n} \\
		\vdots & \ddots & \vdots \\
		x_{m,1} & \cdots & x_{t,n}
	\end{bmatrix}
	& = &
	\begin{bmatrix}
		\begin{bmatrix}\mathbf{u_1} \end{bmatrix} & \hdots & \begin{bmatrix}\mathbf{u_r}\end{bmatrix}
	\end{bmatrix}
	& \cdot &
	\begin{bmatrix}
		 \sigma_1 & \cdots & 0 \\
		\vdots & \ddots & \vdots \\
		0 & \cdots & \sigma_r
	\end{bmatrix}
	& \cdot &
	\begin{bmatrix}
	\begin{bmatrix}\mathbf{v_1}\end{bmatrix} \\ \vdots \\ \begin{bmatrix}\mathbf{v_r}\end{bmatrix}
	\end{bmatrix}
\end{matrix}
\end{equation*}

To reduce the dimensionality of the matrix, a number $k$ of the singular values are kept, and the rest are discarded.  The number of singular values to keep is arbitrary, but implementations almost always keep large singular values ($\sigma_i > 3$ or so) and discard small ones ($\sigma_i < 0.5$).  Intuitively, these larger values are important to the document collection, while smaller ones only serve to contribute to the over-specification.

The product of the resulting matrices $\mathbf{U}_k$, $\mathbf{S}_k$, and $\mathbf{V}^T_k$ produces a dimensionally reduced co-occurrence matrix $\mathbf{X_k}$.
\begin{equation*}
\begin{matrix}
	\mathbf{X_k} & & \mathbf{U}_k & & \mathbf{S}_k & & \mathbf{V}^T_k \\
	\\
	\begin{bmatrix}
		 x_{1,1} & \cdots & x_{1,n} \\
		\vdots & \ddots & \vdots \\
		x_{m,1} & \cdots & x_{m,n}
	\end{bmatrix}
	& = &
	\begin{bmatrix}
		\begin{bmatrix}\mathbf{u_1} \end{bmatrix} & \hdots & \begin{bmatrix}\mathbf{u_k}\end{bmatrix}
	\end{bmatrix}
	& \cdot &
	\begin{bmatrix}
		 \sigma_1 & \cdots & 0 \\
		\vdots & \ddots & \vdots \\
		0 & \cdots & \sigma_k
	\end{bmatrix}
	& \cdot &
	\begin{bmatrix}
	\begin{bmatrix}\mathbf{v_1}\end{bmatrix} \\ \vdots \\ \begin{bmatrix}\mathbf{v_k}\end{bmatrix}
	\end{bmatrix}
\end{matrix}
\end{equation*}

Here, the vectors $\mathbf{u_i} = (u_{i,1},...,u_{i,n})$ and $\mathbf{v_i} = (v_{i,1},...,v_{i,m})$ are left and right singular row vectors for $\mathbf{X}$.

To compare documents and terms in this new latent space, it must be shown that there exists an analog in LSA to the inner space used for finding the similarity in the original vector space model.  The dot products between all documents in the collection is calculated with $\mathbf{X}^T\mathbf{X}$. The following manipulations \cite{Furnas_1988} show that this is equivalent to the following latent space concept:
\begin{equation} \label{eq:innerproductdocument}
    \mathbf{X}^T\mathbf{X} = (\mathbf{USV}^T)^T\mathbf{USV}^T=\mathbf{VSU}^T\mathbf{USV}^T=\mathbf{VSSV}^T=\mathbf{(VS)(VS)}^T
\end{equation}

This means that document comparison is now possible by using the inner products of rows from the $\mathbf{VS}^T$ matrix from equation \ref{eq:innerproductdocument}.

A comparison among terms is done similarly, by first taking:

\begin{equation} \label{eq:innerproductterm}
	\mathbf{X}\mathbf{X}^T = \mathbf{USV}^T(\mathbf{USV}^T)^T=\mathbf{USV}^T\mathbf{VSU}^T=\mathbf{USSU}^T=\mathbf{(US)(US)}^T
\end{equation}

The inner product of rows from equation \ref{eq:innerproductterm}'s $\mathbf{US}$ matrix allow terms to be compared.

Finally, a query is represented as a new document vector $\mathbf{q}$ containing the term counts found in the query.  This can be transformed into the latent space as $\mathbf{q_k} = \mathbf{q}^T\mathbf{U}_k\mathbf{S}_k^{-1}$.  The previously mentioned method for comparing documents in latent space can now be utilized to rank documents.

LSA is a useful technique for improving the quality of query results, but it suffers from a weak mathematical foundation that does not provide a solid set of statistical assumptions about its operations.  As it does not specify any kind of generative model, it produces no clear normalized probability distribution, and instead focuses on finding a rank $k$ matrix that minimizes the Frobenius norm error with the co-occurrence matrix. While using Singular Value Decompositions with limited singular values has been shown \cite{Johnson_1963} to always produce such a rank $k$ matrix, there is not much room to expand the retrieval model to include concepts like query expansion and term dependence.

\subsubsection{Statistical Approaches}

Probabilistic Latent Semantic Analysis (PLSA) is a way of providing a more structured approach to the problem of identifying latent concepts \cite{Hofmann_1999}.  PLSA takes a stronger statistical approach by constructing a generative model for the model.

PLSA represents documents and terms as vectors $\mathbf{D}$ and $\mathbf{W}$, and uses an aspect model that associates an observed class variable $z\in\mathbf{Z}$ with observed documents.  The joint distribution is represented as:
\begin{equation*}
P(d,w)=P(d)\sum_{z\in\mathbf{Z}}P(w|z)P(z|d)
\end{equation*}

The generative model is then fitted through maximum likelihood with the Expectation Maximization (EM) algorithm.

One improvement to PLSA called Latent Dirichlet Allocation (LDA) \cite{Blei_2003} was proposed which seeks to capture more of the document collection's dependence relationships.  Specifically, LDA takes a Bayesian approach and performs inference with prior distributions for terms and documents.  In particular, this method gives more generalization, as it constructs a true generative model that represents both seen and unseen documents.

Both LDA and PLSA reevaluate the mathematical underpinnings of LSA for Information Retrieval, but do so by discarding the linear algebra approach of LSA in favor of a different, more structurally sound statistical model.

\subsubsection{Information Retrieval with Markov Random Fields}

The task of expanding the basic vector space model was approached by \cite{Metzler_2005_MRF} with a formal Markov Random Field framework.  In this approach, three methods were offered for modeling term dependencies: independent, sequential, and fully dependant.  The suggested approach was the sequential dependency graph, containing cliques representing documents, terms, ordered term sequences, and unordered term sequences. The criteria for ranking documents based on sequential dependencies was this ranking function:
\begin{equation} \label{eq:metzlerrank}
	P(D|Q) \propto \sum_{c \in T}{\lambda_Tf_T(c)} + \sum_{c \in O}{\lambda_Of_O(c)} + \sum_{c \in O \cup U}{\lambda_Uf_U(c)}
\end{equation}

The functions $f_T$, $f_O$, and $f_U$ are clique potential functions representing the compatibility of clique in the given distribution. The set of weights $(\lambda_T, \lambda_O, \lambda_U)$ is then learned by using a hill climbing search to optimize the mean average precision.  He showed \cite{Metzler_2005_Max} that the surface is concave, so finding a global maximum is likely.  Clique functions utilize simple smoothing based on a Dirichlet prior to help generalize the term-document space.

This approach uses Markov Random Fields (MRF) as a model for producing the weighted sum of functions relating terms and documents in equation \ref{eq:metzlerrank}.  It is important to note that while, since it is simply another way of stating common information retrieval formulas, this is not by itself a major advance in information retrieval.  Its real value lies instead in the firm foundation that it provides for applying those formulas, as it specifies both the conditional assumptions made by the equations themselves as well as the method for applying them together.  Because it provides such a solid framework for MRF-based document retrieval, its authors successfully build upon this foundation with extensions describing implicit user preference \cite{Metzler_2006_User}, feature selection \cite{Metzler_2007_Feature}, and latent concept expansion \cite{Metzler_2007_Latent}.

\subsection{Overview of MRF Topic Identification}

In order to achieve the level of flexibility and extensibility achieved by \cite{Metzler_2005_MRF} in that MRF model, we propose another MRF that seeks to capture the smoothing gained from the reduced dimensionality co-occurrence matrix in LSA.  A general method for defining MRF will be outlined and applied to a term-document dependency graph. A learning strategy will then demonstrate that LSA's topic clustering can be achieved with the general term-document MRF approach.

\section{Theory} \label{s:theory}

\subsection{Markov Random Fields}

MRFs provide a flexible framework for depicting conditional relationships between a set of random variables.  Unlike similar models such as Markov Chains and Bayesian Networks, MRFs are not limited to specifying one-way (or causal) links between random variables.

\subsubsection{Definition}

MRFs represent a group of random variables with symmetric neighbor relations that satisfy a set \cite{Golden_1996} of conditions:
\begin{itemize}
    \item The probability of any variable given the rest of the MRF is equal to the probability of that variable given its neighbors.
    \item The probability of any set of random variables in the MRF is greater than zero.
\end{itemize}

The first condition, the Markov property for the MRF, means that comparing probabilities is much simpler, since many of the random variables can be ignored when the one being considered does not depend on them.  The second condition simply limits local probabilities to an open interval $(0,1)$.

To obtain a global distribution for random variables in a MRF, it is first necessary to demonstrate the equivalence between the MRF and the Gibbs distribution \cite{Golden_1996}. This can be shown with the Hammersley-Clifford theorem.  This theorem states that given the random vector $\mathbf{x}$, a collection of graph dependencies $\mathcal{G}$ consiting of dependencies based on a symmetric neighbor relation $\omega \subset \mathbf{x} \times \mathbf{x}$, and a set of maximal cliques $C$ on this graph, the random vector is a MRF is given a joint probability distribution:
\begin{equation*}
	P(\mathbf{x}) = \frac{e^{-V(\mathbf{x})}} {\mathcal{Z}}
\end{equation*}
Where $\mathcal{Z}$ here is a normalization constant that is generally infeasible to calculate.  $V(\mathbf{x})$ refers to a family of potential functions that describe the compatibility of clique structures on $\mathbf{x}$.
This equivalence, know as the Hammersley-Clifford theorem, while never published, was proven in later publications \cite{Besag_1974}.

\subsubsection{Constructing an MRF Model}

\textit{Define a Graph Structure}

The first step in constructing a MRF is to produce a graph $\mathcal{G}$ that contains a vector of random variables $\mathbf{x}$ that satisfies the positivity assumption.  This assumption may be restated from its previous definition to say that each random vector may occur with a nonzero probability.  In practice, this constraint is easily met with a well constructed graph.

\textit{Define Clique Structure}

While factorizing the maximal cliques in a given graph has been shown to be NP-complete \cite{Karp_1972}, a well-designed structure can lead to an easily obtainable and semantically meaningful set of cliques.

\textit{Write Clique Potential Functions}

Once clique structures have been defined, it is now necessary to define clique potential functions for them.  These potential functions represent the compatibility of the clique for the particular distribution.

The individual clique potential functions combine as:

\begin{equation} \label{eq:potential_comb}
	V(\mathbf{x}) = \sum_{c \in C}{V^c(\mathbf{x})}
\end{equation}

Where C is a family of clique configurations and $V^c(\mathbf{x})$ refers to the potential function defined for clique configuration $c$.

\textit{Obtain Joint Distribution}

The Hammersley-Clifford Theorem now allows the joint distribution over $\mathbf{x}$ to be defined as:
\begin{equation} \label{eq:hc_joint}
	P(\mathbf{x}) = \frac{e^{-V(\mathbf{x})}} {\mathcal{Z}}
\end{equation}

Applying function \ref{eq:potential_comb} to equation \ref{eq:hc_joint} produces:
\begin{equation} \label{eq:hc_joint_new}
	P(\mathbf{x}) = \frac{e^{-\sum_{c \in C}{V^c(\mathbf{x})}}} {\mathcal{Z}}
\end{equation}

Defining $\mathcal{Z}$ as $\mathcal{Z} = \sum_{y \in S}{e^{-V(\mathbf{y})}}$ where $S$ is the set of all MRF configurations for $\mathbf{x}$, the joint distribution can be written as:
\begin{equation} \label{eq:hc_joint_norm}
	P(\mathbf{x}) = \frac{e^{-\sum_{c \in C}{V^c(\mathbf{x})}}} {\sum_{y \in S}{e^{-V(\mathbf{y})}}}
\end{equation}

\textit{Provide Learning Strategy}

The last step is to define a method for learning MRF parameters.  An example of one learning strategy is the hill climbing approach taken by Metzler to optimize the weights given to the clique potential functions in equation \ref{eq:metzlerrank}.

\subsection{An MRF Model for Information Retrieval}

With these steps defined, it is now possible to construct a MRF model for representing LSA in Information Retrieval.

\subsubsection{Graph Structure}

\begin{figure}
  \centering
    \includegraphics[width=0.5\textwidth]{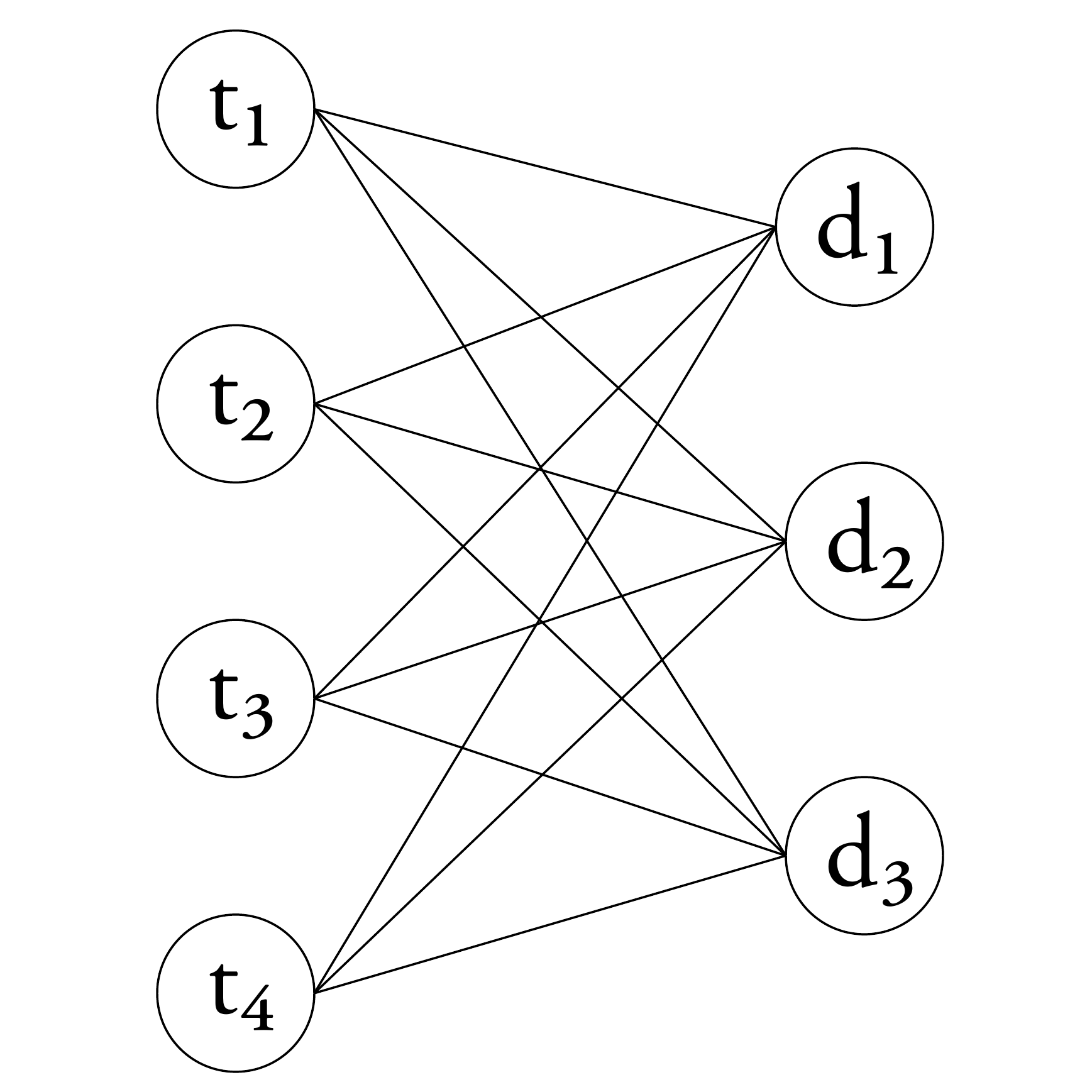}
  \caption{Example Term Document Graph Structure}
  \label{fig:graph_structure}
\end{figure}

The random variables in the MRF will be binary valued random variables.  This choice to declare the random variables as binary-valued leads to the concise clique functions and probability calculation done in \ref{ss:local_probs} and \ref{ss:learning}.

For brevity, it is often convenient to represent the collection of term variables as a row vector $\mathbf{T}$ and the collection of document variables as column vector $\mathbf{D}$.
\begin{align} \label{eq:T_def}
\mathbf{T} &=  \left[t_1,...,t_n\right]\\
\label{eq:D_def}
\mathbf{D} &= \left[d_1,...,d_m\right]^T
\end{align}

Now that the variables in the MRF have been defined, it is necessary to supply neighbor relations $\omega$ on our graph $\mathcal{G}$ representing conditional dependence.  For this graph structure, each document will be connected to every other term, and each term will be connected to every other document.  In this design, the $t$ nodes represent the pool of terms in our collection, while the $d$ nodes represent the documents containing one or more of those terms.

Figure \ref{fig:graph_structure} gives an example of this MRF configuration to visually illustrate the dependence assumptions made in this design.  Semantically, this can be viewed as making the same independence assumptions made in the vector space model that LSA utilizes.  Specificially, we view each document as only being dependent on the terms it contains.  In this way, it is equivalent to the vector-space (bag-of-words model) that stores term counts without any dependence information.

\subsubsection{Clique Definition}

One benefit to the structure we have defined is that it lends itself to easily factored cliques with semantic meaning.  There are three types of cliques in this graph: $C = \{\mathbf{T}, \mathbf{D}, \mathbf{T} \times \mathbf{D}\}$.
Cliques over $\mathbf{T}$ and $\mathbf{D}$ are simple cliques consisting of individual documents and terms, while cliques over $\mathbf{T} \times \mathbf{D}$ are pairs representing term occurrences.

When producing clique functions, the singleton cliques ($\mathbf{T}$ and $\mathbf{D}$) provide an opportunity to weight the importance of terms or documents to the joint distribution.  The pairwise cliques ($\mathbf{T \times D}$) allow the "compatibility" of documents and terms to have an effect on the distribution.

\subsubsection{Clique Potential Functions}
The simplest clique potential function taking the set of random variables $\mathbf{X}$ that may be expressed is the sum of the single and double member clique potential functions:
\begin{equation} \label{eq:basic_pot}
	V(\mathbf{X}) = \sum_i{x_iv_i(x_i)} + \sum_i{\sum_j{x_ix_jv_{ij}(x_i, x_j)}}
\end{equation}

This is just a sum of the single and double member cliques.  One benefit to giving our random variables binary values is that it allows this expression to be simplified greatly without losing any generality.  For any clique whose potential function is $V(x_i) = x_iv_i(x_i)$, it can only take two values: $0$ or $v_i(x_i)$.  Furthermore, if we declare that single clique functions evaluate to members of parameter vectors $\mathbf{b}$ and $\mathbf{g}$ such that $v_i(t_i) = b_i$ and $v_i(d_i) = g_i$, then $t_iv_i(t_i) = 0$ or $b_i$ and $d_iv_i(d_i) = 0$ or $d_i$. Similarly, if potential functions for double member cliques $(t_i, d_j)$ evaluate to members of parameter matrix $\mathbf{W}$ such that $v(t_i, d_j) = W_{ij}$, the expression $t_id_jv_{ij}(t_i, d_j) = 0$ or $W_{ij}$. Given this flexible representation for individual clique potential functions, the sum of all clique potential functions in equation \ref{eq:basic_pot} required for the joint distribution may be written without any loss in generality as:
\begin{equation} \label{eq:sum_pot}
	V(\mathbf{X}) = \sum_i^n{b_it_i} + \sum_j^m{g_jd_j} + \sum_i^n{\sum_j^m{W_{ij}t_id_j}}
\end{equation}

It will occasionally be convenient to notate this function in terms of vectors $\mathbf{T}$ and $\mathbf{D}$ mentioned in equations \ref{eq:T_def} and \ref{eq:D_def}.  This can be done as such:
\begin{equation} \label{eq:sum_vectors}
	V(\mathbf{X}) = \mathbf{bT}^T + \mathbf{gD} + \mathbf{TWD}
\end{equation}

\subsubsection{Joint Distribution}

Now that families of cliques have been defined and given potential functions, an equation for the joint distribution of the MRF model $\mathbf{X}$ may be written, using equation \ref{eq:hc_joint_norm}, as:

\begin{equation} \label{eq:ir_joint}
	P(\mathbf{x}) = \frac
		{\mathrm{exp}(\sum_i^n{b_it_i} + \sum_j^m{g_jd_j} + \sum_i^n{\sum_j^m{W_{ij}t_id_j)}}}
		{\sum_{y \in S}{\mathrm{exp}(\sum_i^n{b_it_i} + \sum_j^m{g_jd_j} + \sum_i^n{\sum_j^m{W_{ij}t_id_j)}}}}
\end{equation}

\subsubsection{Local Probabilities} \label{ss:local_probs}

For information retrieval, local probabilities for individual random variables must be defined.  In particular, this is necessary to find the probability of a particular document $d_i$ given a set of query terms.  For the manipulations required to demonstrate the derivation of this probability, some compact notations will be adopted for the sake of brevity and clarity.
\begin{itemize}
\item The expression $P(X_i=1)$ denotes the probability of some binary variable, either $t_i$ or $d_i$, taking on the value 1.
\item The expression $P(X_{-i})$ denotes the probability of every value in $\mathbf{X}$ except for $X_i$, or $P(X_1,...,X_{i-1},X_{i+1},...,X_d)$.
\item The expression $P(X^{i=k})$ denotes the joint probability of $\mathbf{X}$ such that $X_i=k$, or $P(X_1,...,X_i=k,...,X_d)$.
\end{itemize}

The desired probability may be stated as:
\begin{equation*}
	P(d_i=1|\mathbf{X}_{-i})
\end{equation*}

More clearly, this is equivalent to:
\begin{equation*}
	P(d_i=1|t_1,...,t_n,d_1,...,d_{i-1},d_{i+1},...,d_m)
\end{equation*}

To begin obtaining this probability, it must first be rewritten using a more general form with the compact notation provided above as:
\begin{equation*}
P(X_i=1|X_{-i})
\end{equation*}

This can be manipulated with the following steps:
\begin{align*}
P(X_i=1|X_{-i}) &= \frac{P(X_i=1,X_{-i})}{X_{-1}}\\
&= \frac{P(X^{i=1})}{P(X_{-i})}\\
&= \frac{P(X^{i=1})}{P(X^{i=1})+P(X^{i=0}))}\\
&= \frac{1}{1 + \frac{P(X^{i=0})}{P(X^{i=1})}}
\end{align*}

When the joint probability (equation \ref{eq:hc_joint}) is plugged in, the $\mathcal{Z}$ normalization constants cancel to give:
\begin{align*}
\frac{1}{1 + \frac{P(X^{i=0})}{P(X^{i=1})}} &= \frac{1}{1 + \frac{\mathrm{exp}(-V(X^{i=0}))}{\mathrm{exp}(-V(X^{i=1})))}}\\
&= \frac{1}{1+\mathrm{exp}(-[V(X_{i=1}) - V(X_{i=0})])}
\end{align*}

This takes the form of the sigmoid function, $\varsigma(t) = \frac{1}{1+e^{-t}}$.  It can be written thus as:
\begin{equation} \label{eq:gen_sigmoid}
	P(X_i=1|X_{-i}) = \varsigma(V(X_{i=1}) - V(X_{i=0}))
\end{equation}

In order to write $V(X_{i=0}) - V(X_{i=1})$ in terms of individual random variables and parameters, it is necessary to make several observations about the potential functions.  Because, when $X_i = 0$, the $X_i$ value and its associated parameter will have no contribution to the sum in its family's clique potential function as written in equation \ref{eq:sum_pot}.  It can therefore be written, in the special case considered here in which $X_i$ is a document variable:
\begin{equation} \label{eq:vx1}
	V(X_{i=0}) = \sum_{n}{b_nt_n} + \sum_{m \neq i}{g_md_m} + \sum_n{\sum_{m \neq i}{W_{nm}t_nd_m}}
\end{equation}

Likewise, it is always the case when $X_i = 1$ and $X_i$ is a document variable, that the clique potential function for that MRF is:
\begin{equation} \label{eq:vx0}
	V(X_{i=1}) = \sum_{n}{b_nt_n} + g_i + \sum_{m \neq i}{g_md_m} + \sum_n{W_{ni}t_nd_i} + \sum_n{\sum_{m \neq i}{W_{nm}t_nd_m}}
\end{equation}

When these are plugged into equation \ref{eq:gen_sigmoid}, the shared terms cancel to produce the desired probability in terms of variables and parameters:
\begin{equation} \label{eq:sigmoid_prob_sum}
	P(D_i = 1|X_{-i}) = \varsigma(g_i + \sum_{l=1}^n{W_{li}t_l})
\end{equation}

This can be represented more concisely using vectors as:
\begin{equation} \label{eq:sigmoid_prob_vect}
	P(D_i = 1|X_{-i}) = \varsigma(g_i + \mathbf{W}_i^T\mathbf{T}^T)
\end{equation}

Where $\mathbf{W}_i^T$ indicates the transpose of the $i^{\mathrm{th}}$ column vector of parameter matrix $\mathbf{W}$.

\subsubsection{Learning} \label{ss:learning}

The data that will be used to train the model's parameters will be a set of observation vectors $\mathbf{\hat{T}}^1,...,\mathbf{\hat{T}}^n$ that represent occurrence vectors from the data collection.  $\mathbf{\hat{T}}^i_j$ may indicate the number of times that term $j$ is present in document $1$, but normalized counts such as $tf-idf$ vectors are frequently preferable.

Let us also define a matrix $\mathbf{\hat{T}} =
\begin{bmatrix}
	\begin{pmatrix}\mathbf{\hat{T}}^1\\1\end{pmatrix}, \cdots, \begin{pmatrix}\mathbf{\hat{T}}^n\\1\end{pmatrix}
\end{bmatrix}$
that represents the co-occurence matrix with a row of 1s appended to the bottom.  This can be viewed as a global term that is always on which will be used to estimate parameter $\mathbf{g}$.

The approach for learning parameters will be the maximization of the following sum squared error objective function:

\begin{equation}
	\ell(\mathbf{W}, \mathbf{g}) = \vectornorm{\mathbf{I} - \begin{bmatrix}\mathbf{W}&\mathbf{g}\end{bmatrix}\mathbf{\hat{T}}}_F
\end{equation}

Where $\vectornorm{\mathbf{X}}_F$ indicates the Frobenius norm of some matrix $\mathbf{X}$, and $\mathbf{I}$ is an $n$-dimensional identity matrix whose row vectors represent a configuration of the MRF such that the term variable $T_i$ corresponding with observed occurrence vector $\mathbf{\hat{T}}^i$ is set to 1.

The method of maximizing this will be to solve the following equation:
\begin{equation*}
\mathbf{I} = \begin{bmatrix}\mathbf{W}&\mathbf{g}\end{bmatrix}\mathbf{\hat{T}}
\end{equation*}

The solution is obtained as:
\begin{equation} \label{eq:learning_solution}
\begin{bmatrix}\mathbf{W}&\mathbf{g}\end{bmatrix} = \mathbf{\hat{T}}^\dagger_k
\end{equation}

The term $\mathbf{\hat{T}}^\dagger$ denotes the Moore-Penrose pseudoinverse of matrix $\mathbf{\hat{T}}$.  The expression $\mathbf{\hat{T}}^\dagger_k$ can be calculated by using a singular value decomposition of $\mathbf{\hat{T}}$ keeping $k$ singular values. To obtain matrices $\mathbf{U}_k$, $\mathbf{S}_k$, and $\mathbf{V}_k$.  The pseudo-inverse may now be calculated as:
\begin{equation} \label{eq:defpseudo}
	\mathbf{\hat{T}}^\dagger_k = \mathbf{V}_k\mathbf{S}^{-1}_k\mathbf{U}^T_k
\end{equation}

It is at this point that the comparison to LSA's rank reduction can be drawn.  In this context, the row vectors of the $\begin{bmatrix}\mathbf{W}&\mathbf{g}\end{bmatrix}$ parameter span a $k$-dimensional subspace where $k$ is the number of singular values that have not been set to zero by the SVD operation. It can be shown \cite{Johnson_1963} that this procedure results in finding the $\begin{bmatrix}\mathbf{W}&\mathbf{g}\end{bmatrix}$ that minimizes the sum squared objective function that predicts $\mathbf{I}$ from $\mathbf{\hat{T}}$ using the formula $\mathbf{I} = \begin{bmatrix}\mathbf{W}&\mathbf{g}\end{bmatrix}\mathbf{\hat{T}}$ subject to the constraint that $\begin{bmatrix}\mathbf{W}&\mathbf{g}\end{bmatrix}$ has rank $k$.  This means that the subspace spanned by the row vectors of $\begin{bmatrix}\mathbf{W}&\mathbf{g}\end{bmatrix}$ is reduced in dimensionality in the same way the latent space used to compare documents in LSA is reduced.

\section{Experiments}

\subsection{Method}

The goal of these experiments is to validate the novel approach we have described by comparing its performance to popular retrieval methods.  In particular, we will be looking at various information retrieval metrics and comparing them for varying numbers of singular values taken to reduce the LSA co-occurence matrix or solve \ref{eq:learning_solution} for the MRF approach. In addition, simple vector space term matching will be used as a baseline to evaluate the contribution of term generalization to the algorithms' performance. Since the most obvious algorithm with which to compare our MRF model is the popular Latent Semantic Analysis approach described in \ref{ss:lsa}, it will provide a good baseline for term generalization.

\subsubsection{Data Sets}

The text collections chosen for this paper are the four widely used collections that, together, comprise the \textit{Classic4} data set.  The four collections comprising \textit{Classic4} are:
\begin{itemize}
    \item CRAN - 3204 abstracts from the Cranfield Institute of Technology
    \item CACM - 1460 abstracts from the CACM Journal
    \item CISI - 1460 abstracts from the Institute for Scientific Information
    \item MED - 1033 abstracts from the National Library of Medicine
\end{itemize}

Each collection comes with a set of queries and relevance judgments. This data set was selected based on the quality of the text and query information given as well as its contents.  Academic abstracts would seem to be excellent targets for topic generalization because effective topic generalization manages to resolve the differing jargon that is used in similar academic fields.  This particular data set has also been extensively studied in the past for similar document retrieval approaches such as LSA \cite{Deerwester_1990}, \cite{Hofmann_1999}.

\subsubsection{Procedure}

\textit{Document Collection}

The document collection on which experiments were performed was a combined dataset of the four \textit{Classic4} document collections.  Short terms (below 3 characters), as well as common terms (appearing in 95\% or more documents) were excluded.  Stemming was done with the popular Porter's stemming algorithm \cite{Porter_1997}.

\textit{Vector Space Model}

The simplest baseline for experimentation is done with simple \textit{tf-idf} term matching using vector space methods.  Documents are ranked based on their angular difference from queries in document-term vector space.  The method used involved ranking by highest cosine of the angle, using equation \ref{eq:cos_similarity} given during the description of this approach previously.

\textit{Latent Semantic Analysis}

Document ranking with LSA follows the procedure outlined in section \ref{ss:lsa}.  Specifically, the data collection was loaded as a term-document matrix with \textit{tf-idf} adjustments.  Then a singular value decomposition was done, $\mathbf{X} = \mathbf{USV}^T$, where $\mathbf{X}$ is the co-occurence matrix. Each query $\mathbf{q}_i$ was mapped into the latent space query $\mathbf{L}_i$ as $\mathbf{L}_i = \mathbf{q}_i\mathbf{VS}^{-1}$.  Comparisons with the document collection for query $k$ were then done by finding the maximum cosine angle between latent document $\mathbf{V}_i$ and latent query $\mathbf{L}_k$ for each document $i$.  This can be calculated as: $$\mathrm{cos}(\theta) = \frac{\mathbf{V}_i\cdot\mathbf{L}_k}{\vectornorm{\mathbf{V}}\cdot\vectornorm{\mathbf{L}_k}}$$

The role of the number of values kept from the singular value decomposition is first tested by finding the ideal number of values to keep when decomposing the co-occurence matrix.  Since the style of queries for each collection differs somewhat (samples are given in Appendix \ref{ch:sample_queries}), it is necessary to view the different mean average precision values for each collection's queries.

After this, precision-recall graphs are made using average precision over the set of all queries for each individual document collection.

\textit{Markov Random Field Model}

Document ranking was done by loading the data collection as a term-document matrix with \textit{tf-idf} adjustments and then applying the methods described in part \ref{s:theory} of this paper to obtain the parameters of the MRF.  No weighting is done, the co-occurrence matrix simply records term counts.  The formula used in equation is then used to obtain the probability of a certain document given the terms of the MRF, which are set to match the sample queries given with the collections.

The role of the singular value decomposition is first tested by finding the ideal number of singular values to keep when learning MRF parameters using a method similar to the previous LSA experiment using mean average precision for each collection's set of queries.

Once this is done, it is possible to select good singular value counts for each query collection and create precision-recall graphs based on the average precision values for each set of queries.

\subsection{Results}

\begin{figure}[H]
  \centering
    \includegraphics[width=0.6\textwidth]{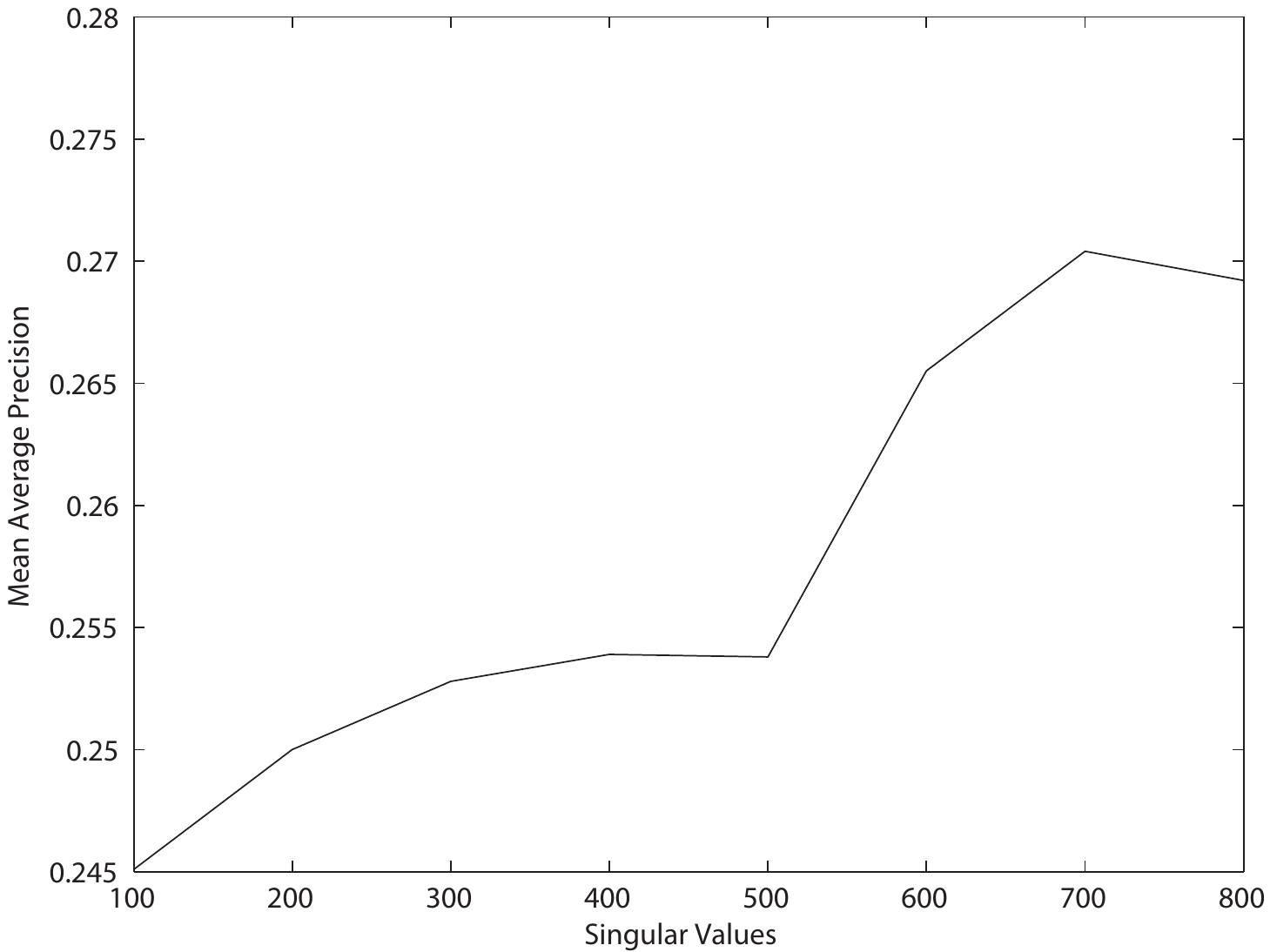}
  \caption{CACM Collection - Mean Precision for Varying Numbers of Singular Values Used (LSA)}
  \label{fig:cacm_map_lsa}
\end{figure}

\begin{figure}[H]
  \centering
    \includegraphics[width=0.6\textwidth]{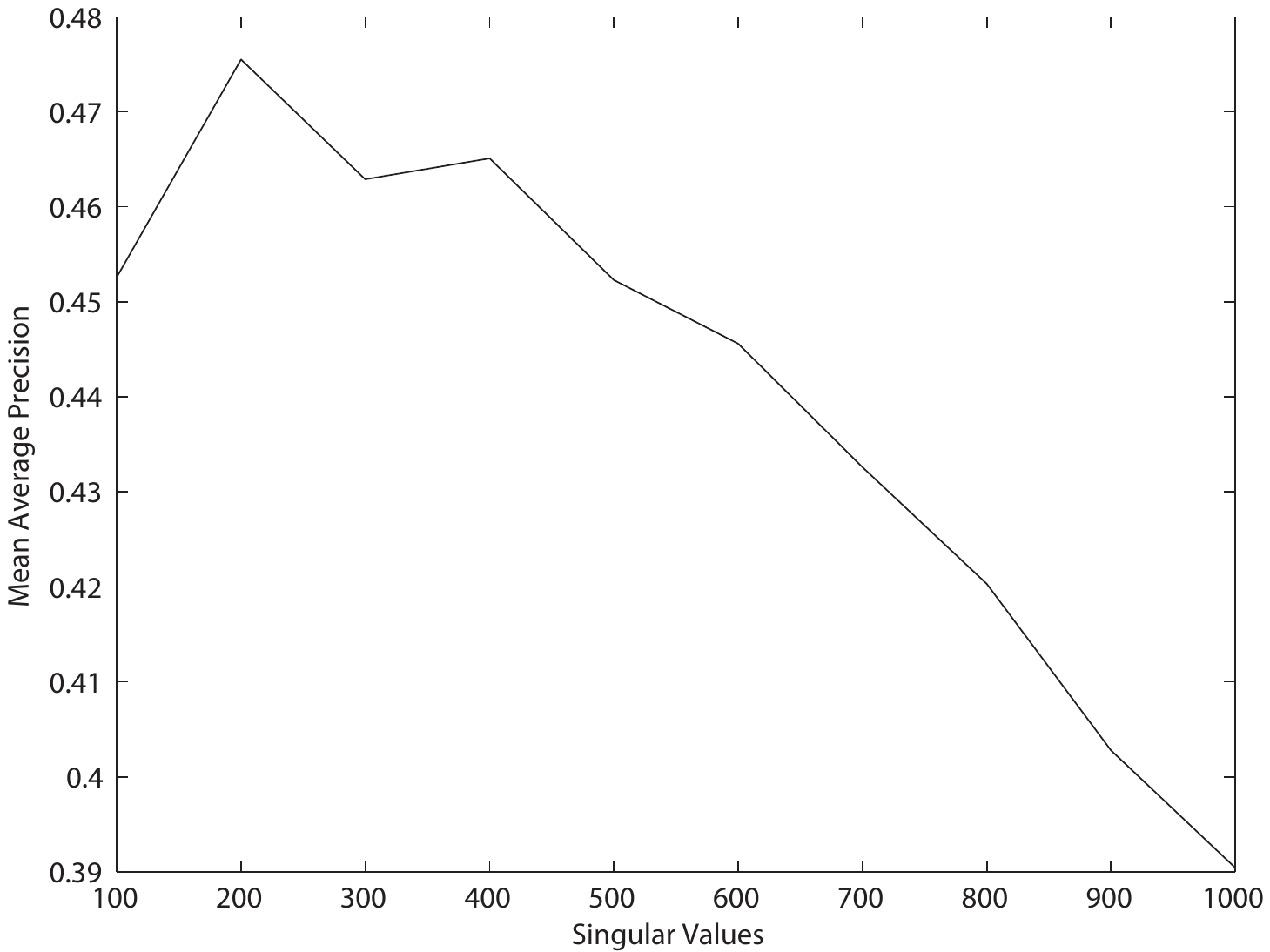}
  \caption{MED Collection - Mean Precision for Varying Numbers of Singular Values Used (MRF)}
  \label{fig:med_map}
\end{figure}

The results for the mean precision versus singular values taken tests (for both LSA and MRF model forms of rank reduction) is shown in Figures \ref{fig:med_map} through \ref{fig:cacm_map_lsa} for the four text collections. Due to the granularity of the mean average precision value difference between differing values kept as well as the large difference between mean average precision values across document collections, each document collection's graph will be shown indepedently.

\begin{figure}[H]
  \centering
    \includegraphics[width=0.6\textwidth]{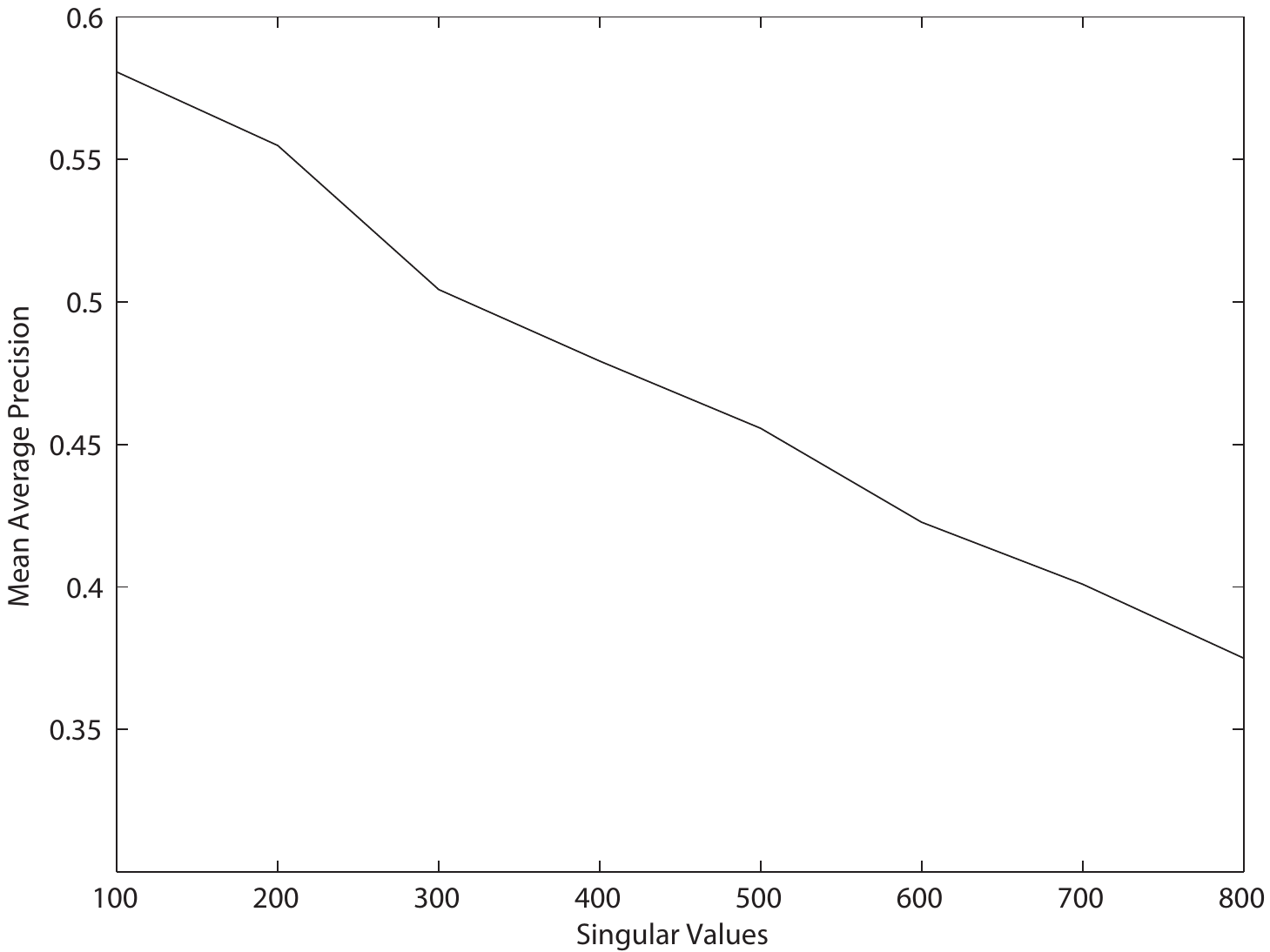}
  \caption{MED Collection - Mean Precision for Varying Numbers of Singular Values Used (LSA)}
  \label{fig:med_map_lsa}
\end{figure}

\begin{figure}[H]
  \centering
    \includegraphics[width=0.6\textwidth]{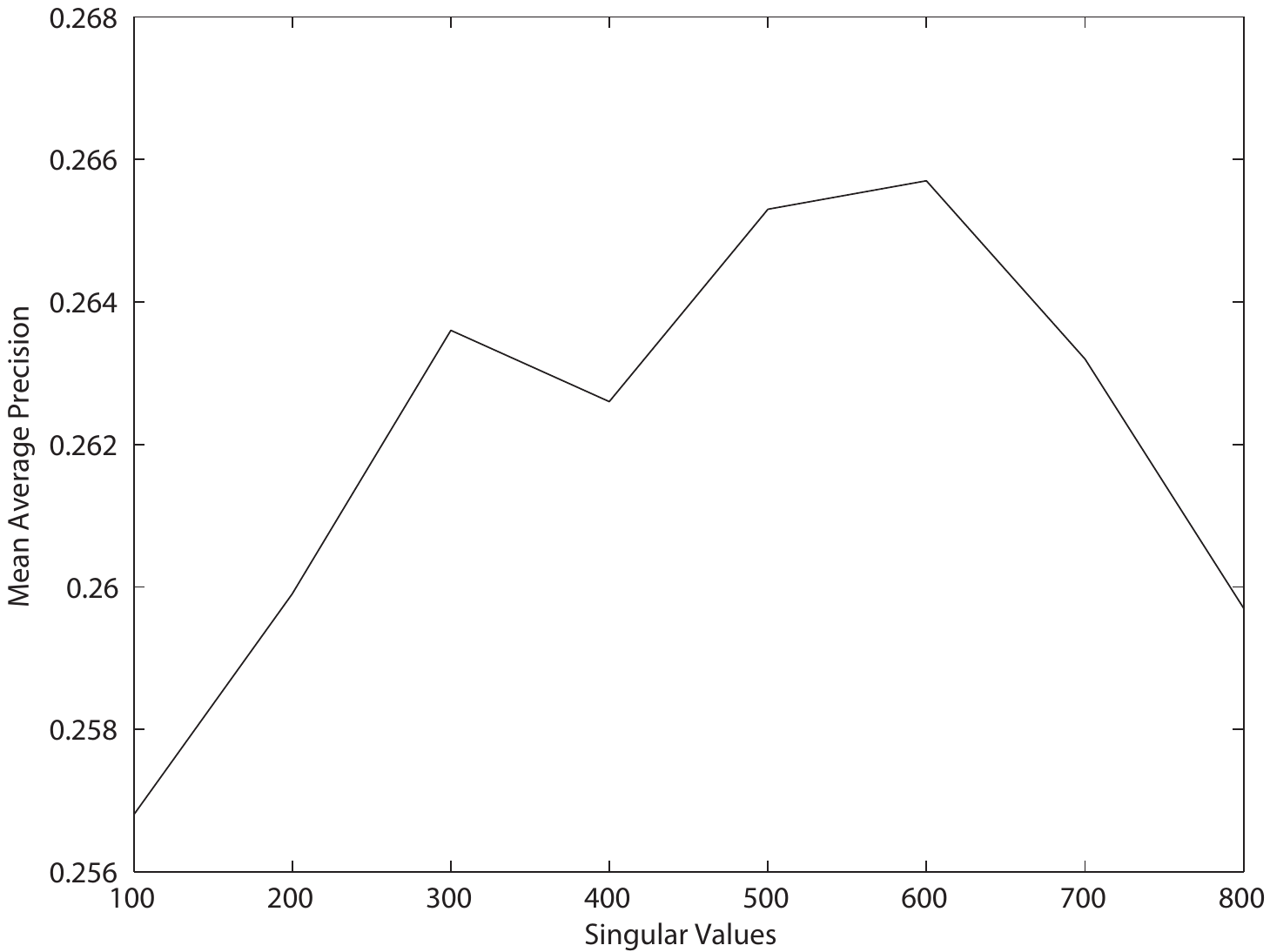}
  \caption{CRAN Collection - Mean Precision for Varying Numbers of Singular Values Used (LSA)}
  \label{fig:cran_map_lsa}
\end{figure}

\begin{figure}[H]
  \centering
    \includegraphics[width=0.6\textwidth]{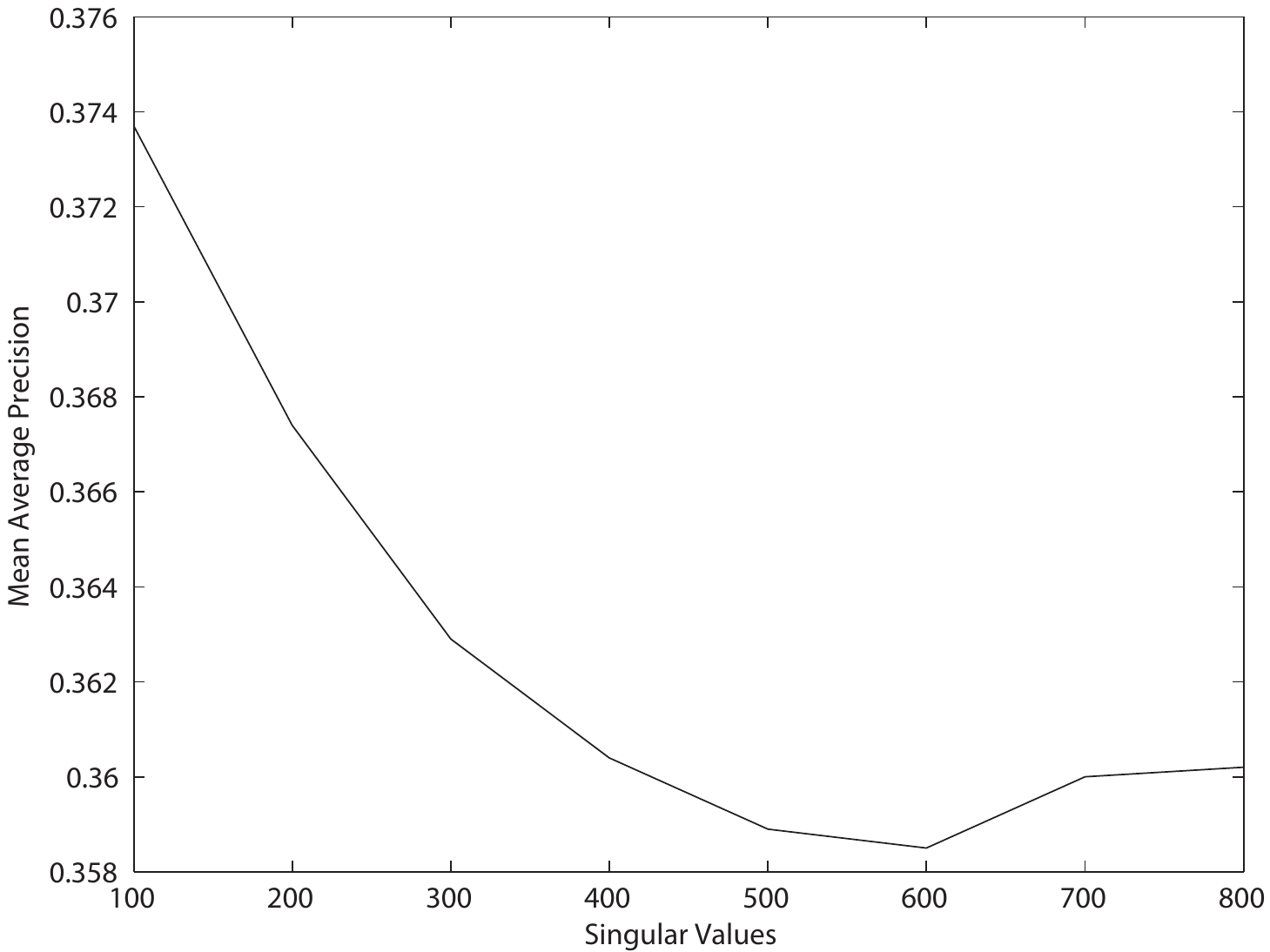}
  \caption{CISI Collection - Mean Precision for Varying Numbers of Singular Values Used (LSA)}
  \label{fig:cisi_map_lsa}
\end{figure}

\begin{figure}[H]
  \centering
    \includegraphics[width=0.6\textwidth]{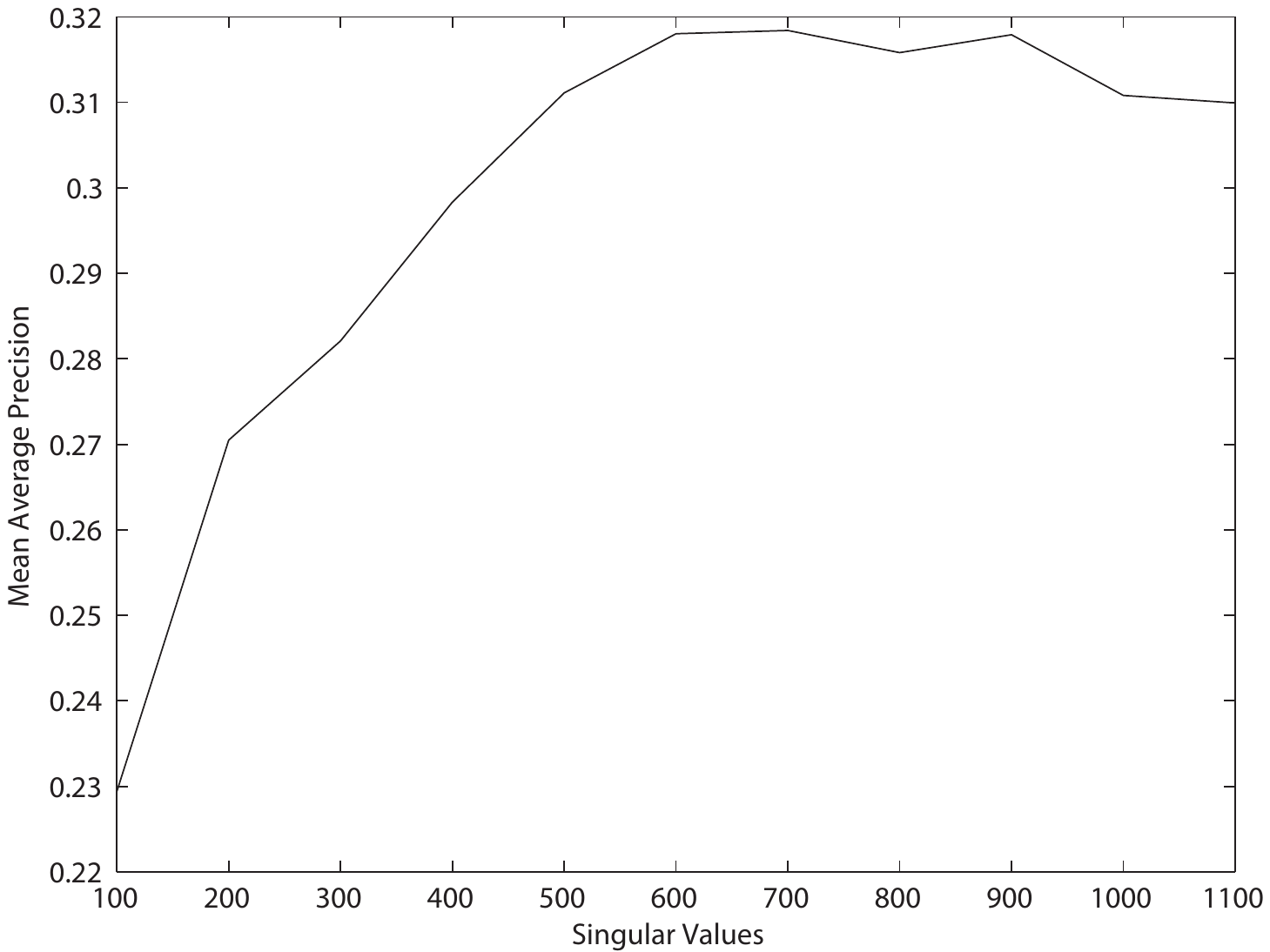}
  \caption{CRAN Collection - Mean Precision for Varying Numbers of Singular Values Used (MRF)}
  \label{fig:cran_map}
\end{figure}

\begin{figure}[H]
  \centering
    \includegraphics[width=0.6\textwidth]{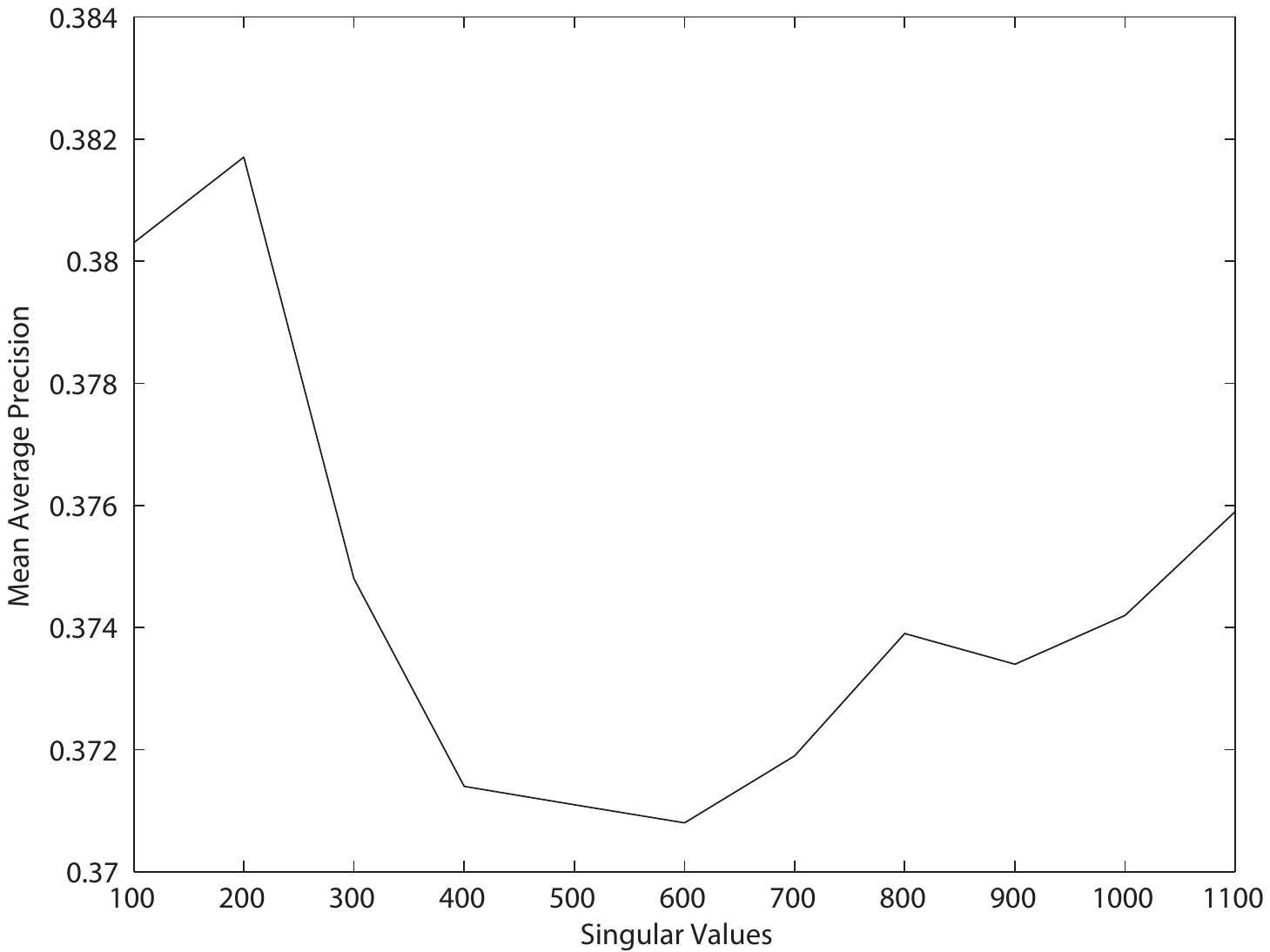}
  \caption{CISI Collection - Mean Precision for Varying Numbers of Singular Values Used (MRF)}
  \label{fig:cisi_map}
\end{figure}

\begin{figure}[H]
  \centering
    \includegraphics[width=0.6\textwidth]{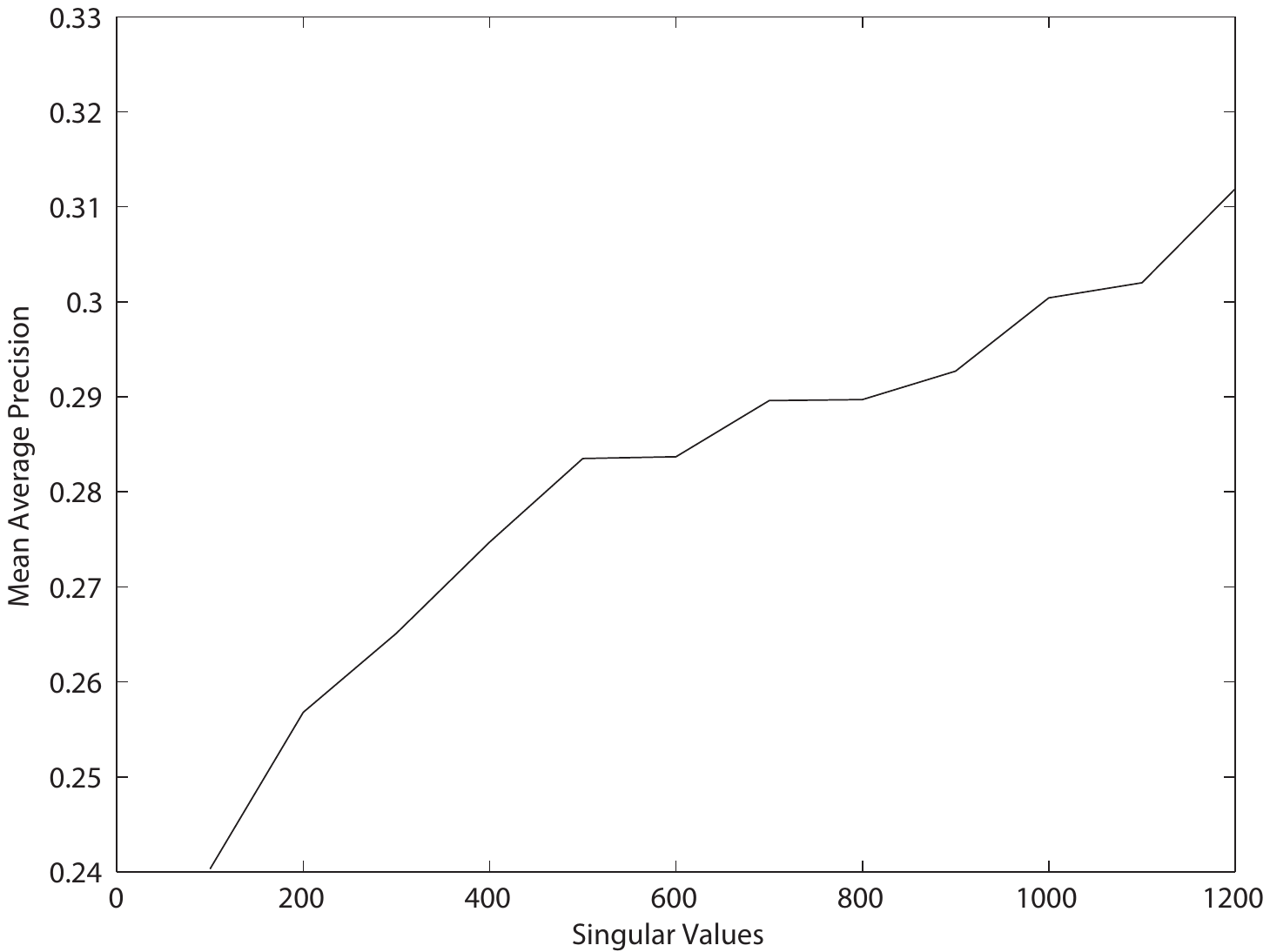}
  \caption{CACM Collection - Mean Precision for Varying Numbers of Singular Values Used (MRF)}
  \label{fig:cacm_map}
\end{figure}

\begin{figure}[H]
  \centering
    \includegraphics[width=0.6\textwidth]{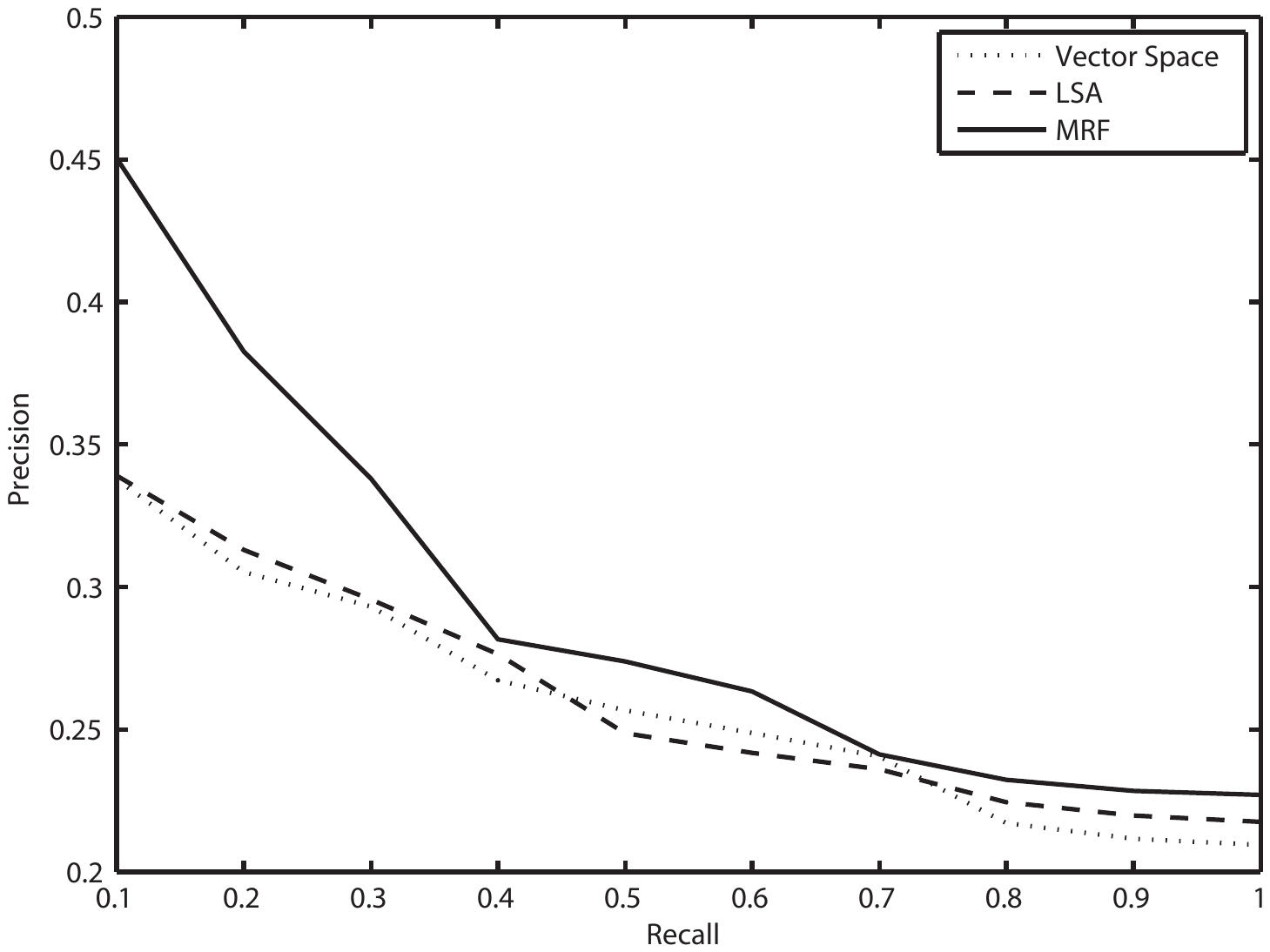}
  \caption{CACM Collection - Precision-Recall}
  \label{fig:cacm_pr}
\end{figure}

\begin{figure}[H]
  \centering
    \includegraphics[width=0.6\textwidth]{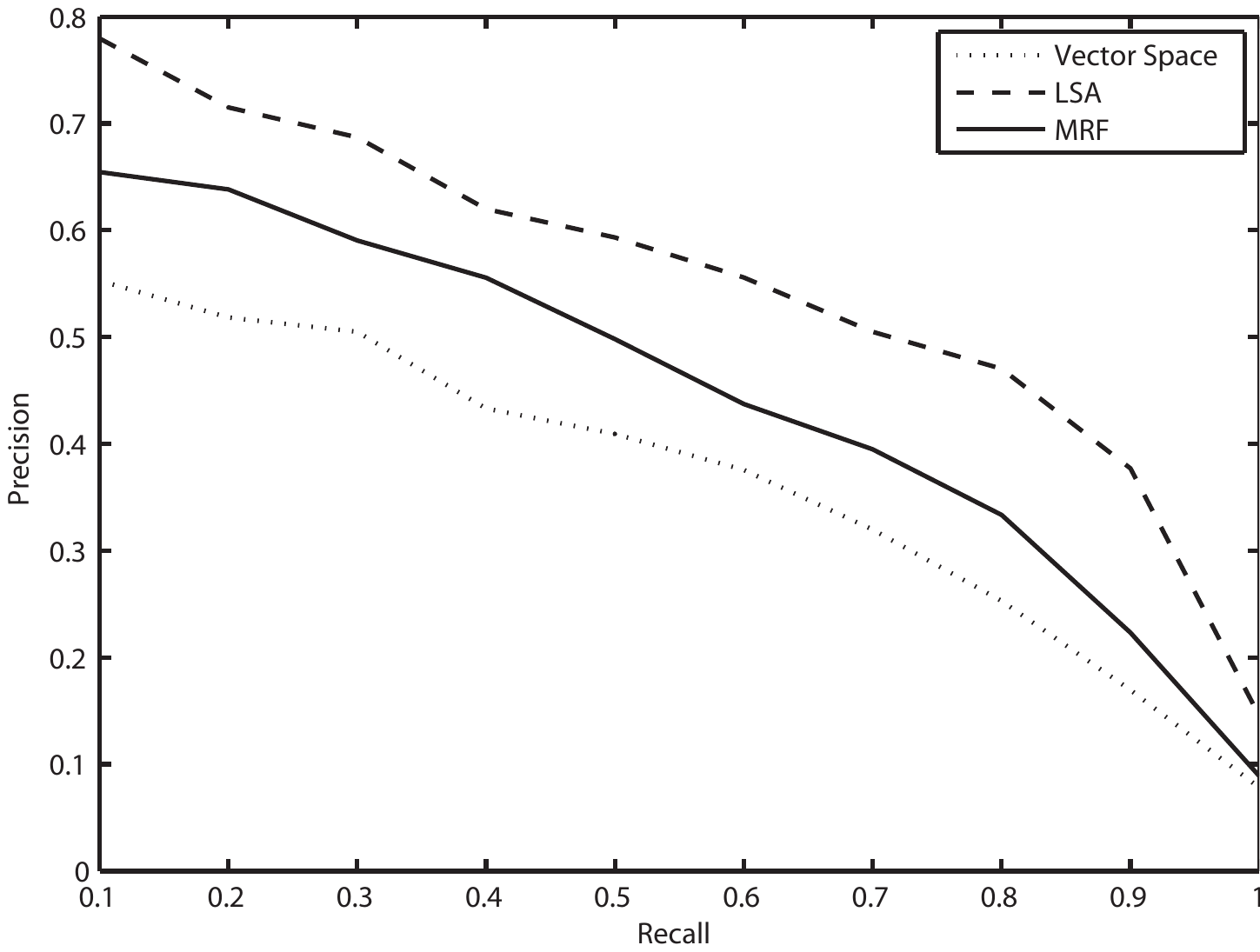}
  \caption{MED Collection - Precision-Recall}
  \label{fig:med_pr}
\end{figure}

Precision-recall graphs for the four collection queries, each using the best number of singular values found in the previous step are given in Figures \ref{fig:med_pr} through \ref{fig:cacm_pr}.  Each graph shows results for vector space indexing, LSA, and MRF retrieval.

\begin{figure}[H]
  \centering
    \includegraphics[width=0.6\textwidth]{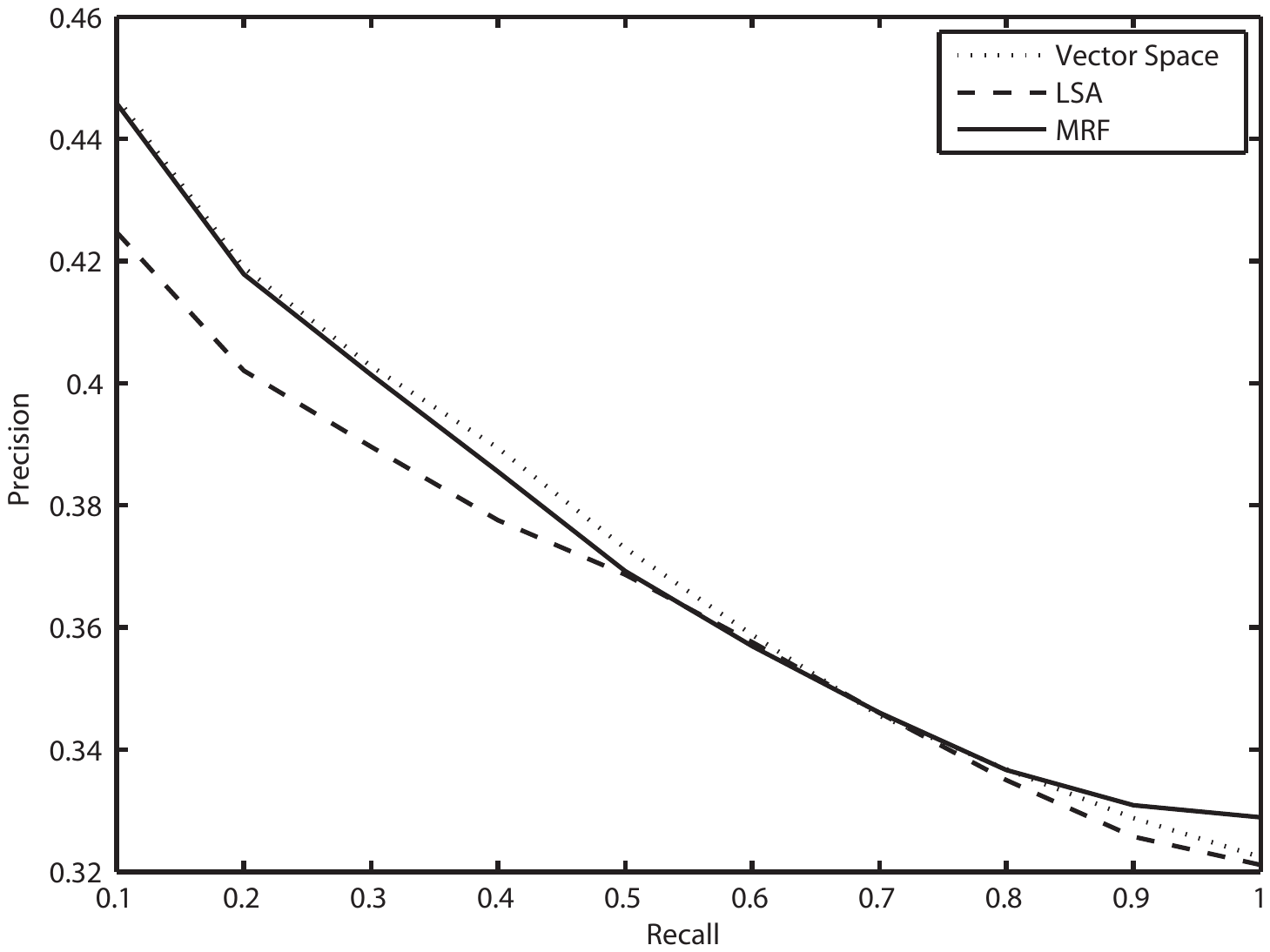}
  \caption{CISI Collection - Precision-Recall}
  \label{fig:cisi_pr}
\end{figure}

\begin{figure}[H]
  \centering
    \includegraphics[width=0.6\textwidth]{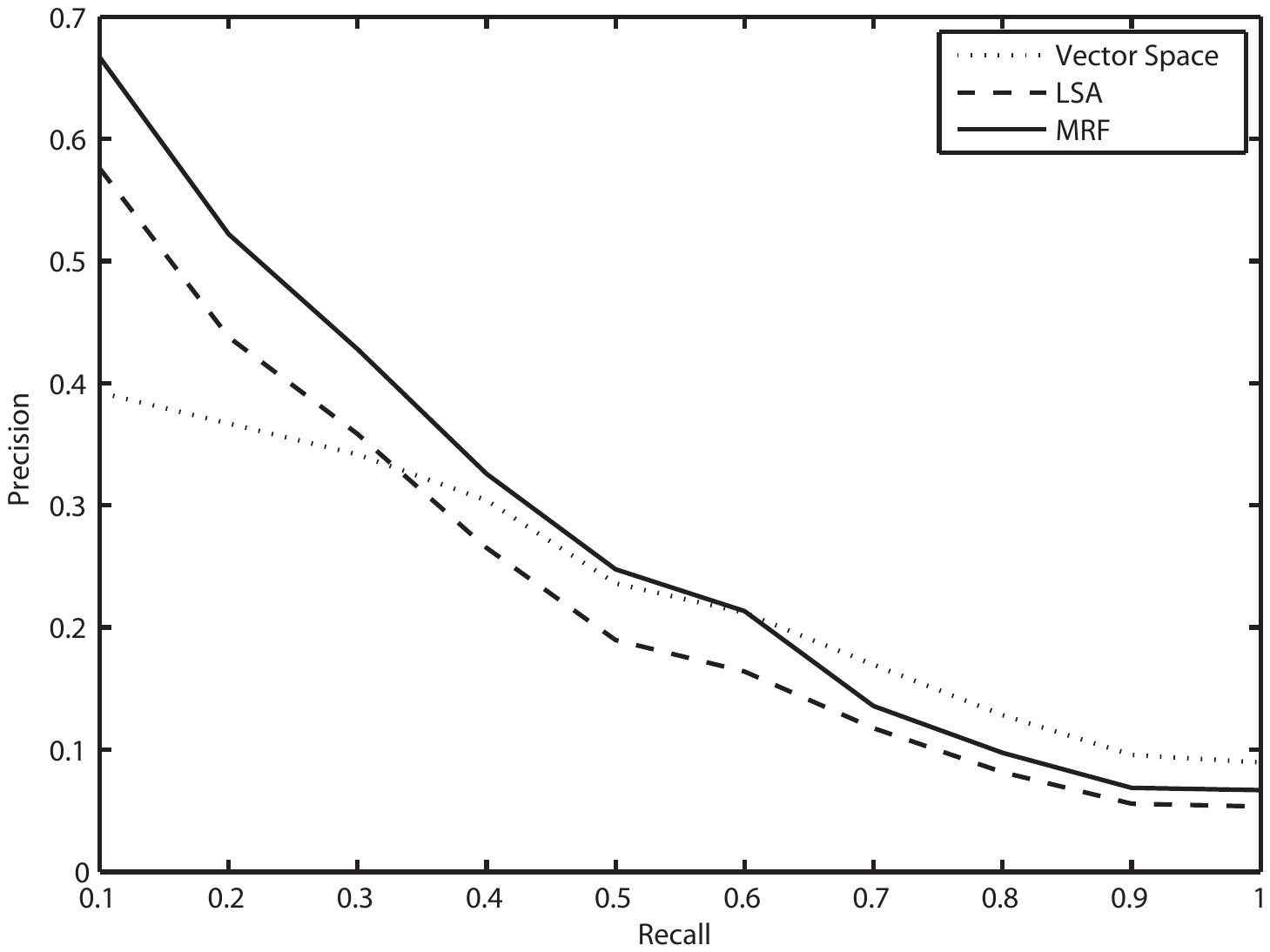}
  \caption{CRAN Collection - Precision-Recall}
  \label{fig:cran_pr}
\end{figure}

\begin{figure}[H]
  \centering
    \includegraphics[width=0.6\textwidth]{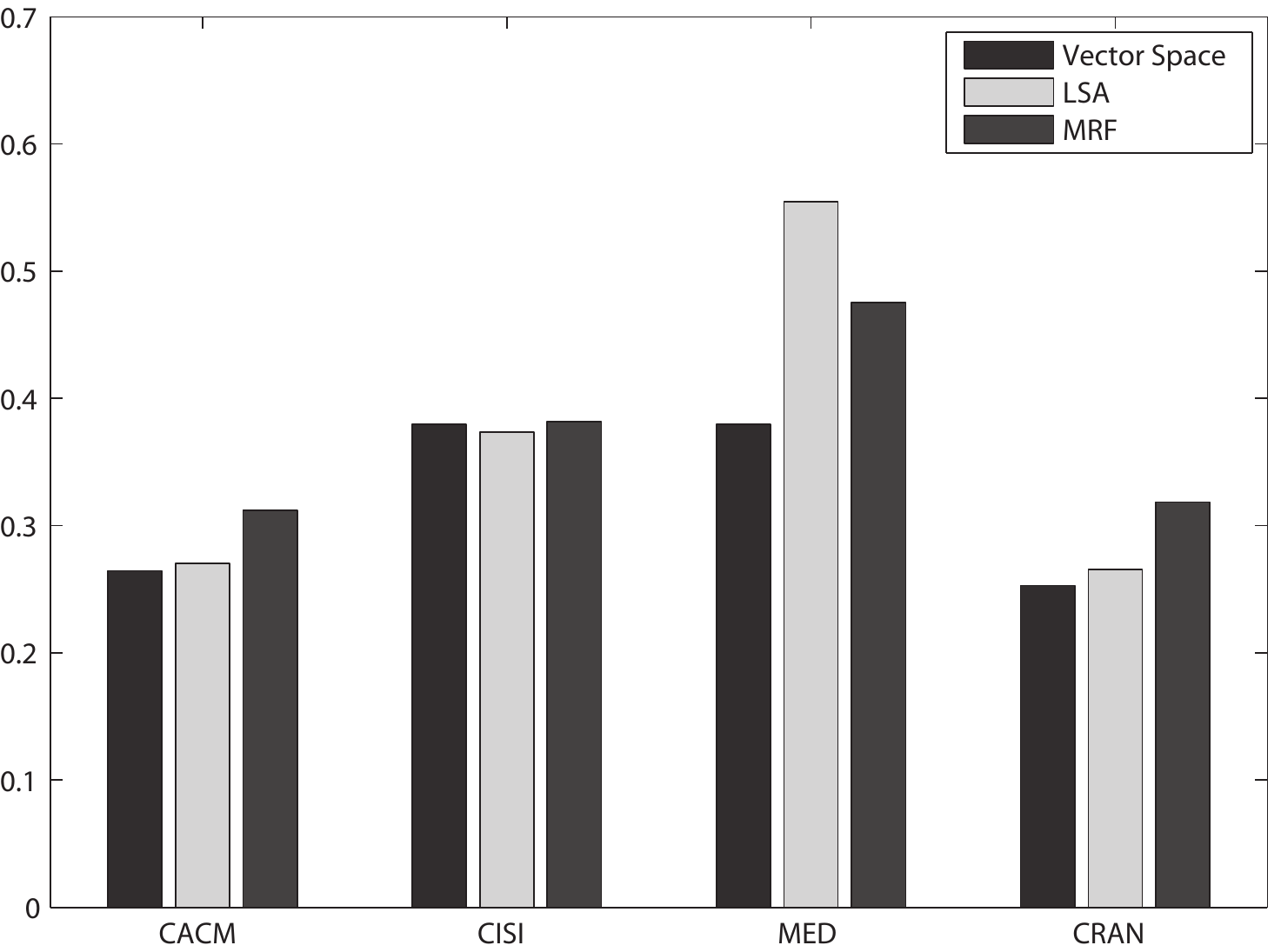}
  \caption{Mean Average Precision Scores for the Three Approaches}
  \label{fig:bar_maps}
\end{figure}

A visual depiction of the mean average precision for each algorithm is shown in figure \ref{fig:bar_maps}.

\subsection{Discussion}

The first experimental result concerned the selection of optimal numbers of singular values for use in the rank reduction in LSA and the pseudo-inverse using the MRF method.  

For LSA, best singular value counts of 100, 600, 100, and 700 were found for the MED, CRAN, CISI, and CACM collections respectively.  It was clear that some collections (MED and CISI) benefitted from smaller counts, while it took much larger counts for CRAN and CACM.  However, these are still fractions of the almost 6000 terms in the original data set. For the MRF model, it seems that certain data sets were better suited to this method than others.  The MED and CISI had maximums at low (200) singular values.  CRAN took 900 singular values before tapering off in performance.  CACM did not seem suited to the reduced dimensionality, as it continued to increase in performance after reaching around a fifth of possible singular values (1200 out of 5896).

Precision-recall graphs show promise in the MRF method.  It succeeds remarkably in querying CISI, where LSA has been known to show significantly worse performance than simple vector space methods \cite{Deerwester_1990}.  For the CACM and CRAN collections, it outperformed LSA and either matched or outperformend vector space methods.  The only collection in which LSA was strictly superior was the MED collection. It is not entirely clear why this is the case, although the MED collection is the smallest of the collections and has a very small query collection, so it is possible that some aspect of this unusual collection produced such polarized results.  However, even in this case, the MRF method still outperformed simple vector space methods.

Not only do these results suggest that our approach is sound for information retrieval, but they also give credibility to our previous assertion that the benefits from rank reduction in LSA can be matched by reducing the dimensionality of the MRF parameter matrix $\mathbf{W}$.

\section{Conclusion}

\subsection{Summary}

\subsubsection{Theory}

In this paper, we have presented a methodical approach to defining a Markov Random Field (MRF) that captures the independence assumptions made in document indexing with Latent Semantic Analysis (LSA).  A clearly defined graph structure produces a set of semantically meaningful clique potential functions describing the compatibility of documents, terms, and document-term pairs in the model.

After declaring these properties of our graph, we utilized the Hammersley-Clifford theorem to state that the joint distribution of the random variables in our graph is defined with a Gibbs distribution.  Some manipulation of probabilities was done to find a concise expression for the probability of any particular document given a set of terms.

Finally, a method for learning parameters was proposed.  This method minimizes a sum squared error using the Moore-Penrose pseudoinverse.  Because this pseudoinverse relies on a singular value decomposition to produce the desired parameters, it is possible to limit the singular values does and achieve the same benefits as the rank reduction in LSA.

\subsubsection{Results}

Experiments were carried out on the medium-sized \textit{Classic4} data set of scientific abstracts.  The results showed that, like LSA, the number of singular values kept in the rank reduction affects performance.  Once the largest effective number of singular values for each collection was determined, queries for each collection were executing on an MRF formed by learning with that number of singular values.  Average precision-recall graphs for the MRF approach as well as the LSA and vector space methods were constructed for each set of queries that showed effective retrieval by the MRF method.

The results of these queries were promising.  Even though CISI was previously described as being difficult, precision scores remained above 0.2 for all recall values.  Both MED and CRAN collections produced excellent results, with 0.4755 and 0.3184 mean average precision scores respectively.  CISI produced a mean average precision of 0.3817, a surprisingly high score for such a difficult collection.  The most difficult collection with this method proved to be CACM with a score of 0.3119, but that is not significantly lower than the others.  LSA was only able to outperform our approach on the small MED collection, but the MRF model outperformed LSA on the other 3 collections. The efficacy of our method as a document retrieval engine for difficult collections is suggested by these results.

\subsection{Uses and Extensions}

The greatest benefit of our approach is its potential for future expansion.  Now that a clear statistical model has been proposed that utilizes rank reduction in a similar manner to LSA, the next step will be to add new assumptions to the MRF model that produce more intelligent results.  Term dependencies, hierarchical document structures, and query expansion are several ideas for future research with this approach.

\appendix
\section{Sample Documents from Classic4 Data Set}
\subsection{CRAN}
\begin{quote}
\texttt{experimental investigation of the aerodynamics of a
wing in a slipstream.
  an experimental study of a wing in a propeller slipstream was
made in order to determine the spanwise distribution of the lift
increase due to slipstream at different angles of attack of the wing
and at different free stream to slipstream velocity ratios.  the
results were intended in part as an evaluation basis for different
theoretical treatments of this problem.
  the comparative span loading curves, together with
supporting evidence, showed that a substantial part of the lift increment
produced by the slipstream was due to a /destalling/ or
boundary-layer-control effect.  the integrated remaining lift
increment, after subtracting this destalling lift, was found to agree
well with a potential flow theory.
  an empirical evaluation of the destalling effects was made for
the specific configuration of the experiment.}
\end{quote}

\subsection{CISI}
\begin{quote}
\texttt{The present study is a history of the DEWEY Decimal
Classification.  The first edition of the DDC was published
in 1876, the eighteenth edition in 1971, and future editions
will continue to appear as needed.  In spite of the DDC's
long and healthy life, however, its full story has never
been told.  There have been biographies of Dewey
that briefly describe his system, but this is the first
attempt to provide a detailed history of the work that
more than any other has spurred the growth of
librarianship in this country and abroad.}
\end{quote}

\subsection{CACM}
\begin{quote}
\texttt{This paper discusses the limited problem of
recognition and retrieval of a given misspelled
name from among a roster of several hundred names, such
as the reservation inventory for a given flight
of a large jet airliner.  A program has been developed
and operated on the Telefile (a stored-program
core and drum memory solid-state computer) which will
retrieve passengers' records successfully, despite
significant misspellings either at original entry time
or at retrieval time.  The procedure involves
an automatic scoring technique which matches the names
in a condensed form. Only those few names most
closely resembling the requested name, with their phone
numbers annexed, are presented for the agents
final manual selecton.  The program has successfully
isolated and retrieved names which were subjected
to a number of unusual (as well as usual) misspellings.}
\end{quote}

\subsection{MED}
\begin{quote}
\texttt{correlation between maternal and fetal plasma levels of glucose and free
fatty acids.
  correlation coefficients have been determined between the levels of
glucose and ffa in maternal and fetal plasma collected at delivery.
significant correlations were obtained between the maternal and fetal
glucose levels and the maternal and fetal ffa levels. from the size of
the correlation coefficients and the slopes of regression lines it
appears that the fetal plasma glucose level at delivery is very strongly
dependent upon the maternal level whereas the fetal ffa level at
delivery is only slightly dependent upon the maternal level.}
\end{quote}

\section{Sample Queries from the Classic4 Data Set} \label{ch:sample_queries}
\subsection{CRAN}
\subsubsection{Query 1 of 365}
\begin{quote}
\texttt{what similarity laws must be obeyed when constructing aeroelastic models
of heated high speed aircraft.}
\end{quote}

\subsubsection{Query 2 of 365}
\begin{quote}
\texttt{what are the structural and aeroelastic problems associated with flight
of high speed aircraft.}
\end{quote}

\subsection{CISI}
\subsubsection{Query 1 of 112}
\begin{quote}
\texttt{What problems and concerns are there in making up descriptive titles?
What difficulties are involved in automatically retrieving articles from
approximate titles?
What is the usual relevance of the content of articles to their titles?}
\end{quote}

\subsubsection{Query 2 of 112}
\begin{quote}
\texttt{How can actually pertinent data, as opposed to references or entire articles
themselves, be retrieved automatically in response to information requests?}
\end{quote}

\subsection{CACM}
\subsubsection{Query 1 of 64}
\begin{quote}
\texttt{What articles exist which deal with TSS (Time Sharing System), an
operating system for IBM computers?}
\end{quote}

\subsubsection{Query 2 of 64}
\begin{quote}
\texttt{I am interested in articles written either by Prieve or Udo Pooch}
\end{quote}

\subsection{MED}

\subsubsection{Query 1 of 30}
\begin{quote}
\texttt{the crystalline lens in vertebrates, including humans.}
\end{quote}

\subsubsection{Query 2 of 30}
\begin{quote}
\texttt{the relationship of blood and cerebrospinal fluid oxygen concentrations
or partial pressures.  a method of interest is polarography.}
\end{quote}

\bibliographystyle{alpha}
\bibliography{mrf_paper_ref}

\begin{thebibliography}{10}

\bibitem{Deerwester_1990}
Scott Deerwester, Susan~T. Dumais, George~W. Furnas, Thomas~K. Landauer, and
  Richard Harshman.
\newblock Indexing by latent semantic analysis.
\newblock {\em Journal of the American Society for Information Science},
  41(6):391--407, 1990.

\bibitem{Furnas_1988}
G.~W. Furnas, S.~Deerwester, S.~T. Dumais, T.~K. Landauer, R.~A. Harshman,
  L.~A. Streeter, and K.~E. Lochbaum.
\newblock Information retrieval using a singular value decomposition model of
  latent semantic structure.
\newblock In {\em Proceedings of the 11th annual international ACM SIGIR
  conference on Research and development in information retrieval}, SIGIR '88,
  pages 465--480, New York, NY, USA, 1988. ACM.

\bibitem{Salton_1975}
G.~Salton, A.~Wong, and C.~S. Yang.
\newblock A vector space model for automatic indexing.
\newblock {\em Commun. ACM}, 18:613--620, November 1975.

\bibitem{Johnson_1963}
Richard Johnson.
\newblock On a theorem stated by {E}ckart and {Y}oung.
\newblock {\em Psychometrika}, 28:259--263, 1963.
\newblock 10.1007/BF02289573.

\bibitem{Hofmann_1999}
Thomas Hofmann.
\newblock Probabilistic latent semantic indexing.
\newblock In {\em Proceedings of the 22nd annual international ACM SIGIR
  conference on Research and development in information retrieval}, SIGIR '99,
  pages 50--57, New York, NY, USA, 1999. ACM.

\bibitem{Blei_2003}
David~M. Blei, Andrew~Y. Ng, and Michael~I. Jordan.
\newblock Latent dirichlet allocation.
\newblock {\em J. Mach. Learn. Res.}, 3:993--1022, March 2003.

\bibitem{Metzler_2005_MRF}
Donald Metzler and W.~Bruce Croft.
\newblock A markov random field model for term dependencies.
\newblock In {\em Proceedings of the 28th annual international ACM SIGIR
  conference on Research and development in information retrieval}, SIGIR '05,
  pages 472--479, New York, NY, USA, 2005. ACM.

\bibitem{Metzler_2005_Max}
Donald Metzler, W.~Bruce Croft, and Andrew Mccallum.
\newblock Direct maximization of rank-based metrics for information retrieval.
\newblock Technical report, 2005.

\bibitem{Metzler_2006_User}
Donald Metzler and W.~Bruce Croft.
\newblock Beyond bags of words: Modeling implicit user preferences in
  information retrieval, 2006.

\bibitem{Metzler_2007_Feature}
Donald Metzler.
\newblock Automatic feature selection in the markov random field model for
  information retrieval.
\newblock In {\em In Proceedings of CIKM'07}, 2007.

\bibitem{Metzler_2007_Latent}
Donald Metzler and W.~Bruce Croft.
\newblock Latent concept expansion using markov random fields.
\newblock In {\em In Proceedings of the 30th annual international ACM SIGIR
  conference on Research and development in information retrieval}, 2007.

\bibitem{Golden_1996}
Richard~M. Golden.
\newblock {\em Mathematical Methods for Neural Network Analysis and Design}.
\newblock MIT Press, Cambridge, MA, USA, 1st edition, 1996.

\bibitem{Besag_1974}
Julian Besag.
\newblock {Spatial Interaction and Statistical-Analysis of Lattice Systems}.
\newblock {\em {Journal of the Royal Statistical Society Series
  B-Methodological}}, {36}({2}):{192--236}, {1974}.

\bibitem{Karp_1972}
Richard~M. Karp.
\newblock Reducibility among combinatorial problems.
\newblock In Michael Jünger, Thomas~M. Liebling, Denis Naddef, George~L.
  Nemhauser, William~R. Pulleyblank, Gerhard Reinelt, Giovanni Rinaldi, and
  Laurence~A. Wolsey, editors, {\em 50 Years of Integer Programming 1958-2008},
  pages 219--241. Springer Berlin Heidelberg, 2010.

\bibitem{Porter_1997}
M.~F. Porter.
\newblock {\em An algorithm for suffix stripping}, pages 313--316.
\newblock Morgan Kaufmann Publishers Inc., San Francisco, CA, USA, 1997.

\end{thebibliography}
\end{document}